\documentclass[%
aip,
amsmath,amssymb,
reprint,%
]{revtex4-1}

\usepackage{graphicx}
\usepackage{dcolumn}
\usepackage{bm}

\usepackage[utf8]{inputenc}
\usepackage[T1]{fontenc}
\usepackage{mathptmx}
\usepackage{gensymb}
\usepackage{float}

\begin{document}

\title{Q-factor optimization for high-beta 650 MHz cavities for PIP-II}

\author{M. Martinello}
\email{mmartine@fnal.gov}
 \affiliation{Fermi National Accelerator Laboratory, Batavia, Illinois 60510, USA.}
      \author{D.J. Bice}%
 \affiliation{Fermi National Accelerator Laboratory, Batavia, Illinois 60510, USA.}
   \author{C. Boffo} 
  \affiliation{Fermi National Accelerator Laboratory, Batavia, Illinois 60510, USA.}
   \author{S.K. Chandrasekeran}%
 \affiliation{Fermi National Accelerator Laboratory, Batavia, Illinois 60510, USA.}
      \author{G.V. Eremeev}%
 \affiliation{Fermi National Accelerator Laboratory, Batavia, Illinois 60510, USA.}
      \author{F. Furuta}%
 \affiliation{Fermi National Accelerator Laboratory, Batavia, Illinois 60510, USA.}
\author{A. Grassellino}%
 \affiliation{Fermi National Accelerator Laboratory, Batavia, Illinois 60510, USA.}
  \author{O. Melnychuk}%
 \affiliation{Fermi National Accelerator Laboratory, Batavia, Illinois 60510, USA.}
   \author{D.A. Sergatskov}%
 \affiliation{Fermi National Accelerator Laboratory, Batavia, Illinois 60510, USA.}
  \author{G. Wu}%
 \affiliation{Fermi National Accelerator Laboratory, Batavia, Illinois 60510, USA.}
     \author{T.C. Reid}%
 \affiliation{Argonne National Laboratory, Argonne, Illinois 60439, USA.}

\date{\today}

\begin{abstract}
 High Q-factors are of utmost importance to minimize losses of superconducting radio-frequency cavities deployed in continuous wave particle accelerators. This study elucidates the surface treatment that can maximize the Q-factors in high-beta 650 MHz elliptical niobium cavities. State-of-the-art surface treatments are applied in many single-cell cavities, and surface resistance studies are performed to understand the microwave dissipation at this unexplored frequency. The nitrogen doping treatment is confirmed to be necessary to maximize the Q-factors at medium RF fields. We applied this treatment in 5-cell high-beta 650 MHz cavities and demonstrated that extremely high Q-factors were obtained at medium RF fields with this treatment. We also demonstrated that adding a cold electropolishing step after N-doping is crucial to push the quench field of multicell cavities to higher gradients.
\end{abstract}

\maketitle

\section{Introduction}
The Proton Improvement Plan II (PIP-II) project aims to upgrade the Fermilab accelerator complex to power the world’s most intense high-energy neutrino beam for the Deep Underground Neutrino Experiment (DUNE). The high-power proton beam will also enable muon-based experiments and a broad physics research program \cite{Merminga}.

PIP-II includes the construction of a 215 m superconducting linear accelerator (linac) that will accelerate protons up to 800 MeV by using 5 different types of superconducting cavities: half-wave resonators (HWR), single-spoke resonators (SSR1 and SSR2), and low- and high-beta 650 MHz elliptical cavities (LB650 and HB650). The last section of the linac will be composed of HB650 cavities, which are 5-cell $\beta = 0.92$ 650 MHz elliptical cavities. These cavities require high Q-factors ($Q_0=3 \times 10^{10}$) at relatively high gradients ($E_{\mathrm{acc}}=18.8$ MV/m, $B_{\mathrm{pk}}=73.1$ mT) to limit the power dissipation, and consequently cryogenic consumption during the operation in a continuous wave mode of the linac \cite{PIP-IIdesign}. 

The specifications for the PIP-II HB650 cavities are very challenging compared to specifications of cavities with similar frequency and geometry implemented in other machines such as the Spallation Neutron Source (SNS) at the Oak Ridge National Laboratory (ORNL) and the European Spallation Source (ESS)\cite{SNS, ESS}.
Exceeding these specifications is not straightforward. Decades of research in the superconducting radio frequency (SRF) field have focused on improving the performance of 1.3 GHz Tesla-type cavities, for which challenging specifications were required by major projects such as the linear coherent light source upgrade, LCLS-II, at the SLAC National Laboratory, and the International Linear Collider (ILC)\cite{LCLS-II, ILC_Jlab}. It is thus unclear how performance can be optimized for 650 MHz cavities.
In addition, previous studies have shown that the Q-factor optimization of these cavities at medium RF fields is particularly challenging, since the field dependence of the temperature-dependent component of the surface resistance, which is often called the BCS surface resistance, is unfavorable at this frequency; it more rapidly increases with the RF field than in higher frequency cavities and does not show reversal behavior after N-doping \cite{Martinello_SRF_2017, Martinello_PRL}. Theoretical studies soon demonstrated that this behavior is due to non-equilibrium effects being negligible in niobium cavities resonating below 1 GHz \cite{Kubo_Noneq}.

The goal of the present article is to clarify which surface treatment is capable of maximizing the performance of high-beta 650 MHz cavities, indicating a possible processing path for HB650 cavities for the PIP-II project. For this purpose, we analyzed the performance of several single-cell niobium $\beta=0.9$ 650 MHz cavities, which were subjected to different state-of-the-art surface treatments (electropolishing, low-temperature baking and N doping). 

The performance of single-cell cavities was then analyzed in detail. In this paper, we present the field dependence of the surface resistance components, high-field Q-slope onset and quench field as a function of different surface treatments. The most promising surface treatment obtained from this study was applied to 5-cell $\beta=0.9$ niobium cavities, and will be applied to 5-cell $\beta=0.92$ cavities to qualify the first set of cavities for the PIP-II prototype cryomodule.
The RF design of the $\beta=0.9$ and $\beta=0.92$ HB650 cavities is very similar and enables the combined use of both cavity types in the prototype cryomodule (pCM). According to the final PIP-II linac design, only $\beta=0.92$ HB650 cavities will be procured during the production phase \cite{Yakovlev_PAC2011, Lunin}. The prototype CM testing will validate the cavity processing protocol, verifying that the PIP-II specifications can be met.

\section{Single-cell cavities studies}

The cavities analyzed in this paper were fabricated using high residual-resistivity-ratio (RRR>300) niobium. The following sections describe the results obtained in clean niobium, low-T baked and N-doped cavities, which underscore the specific processing and surface treatment flow in each case.  All RF measurements were conducted in the cavity vertical test stand (VTS) facility at Fermilab within the temperature range of 1.4-2.0 K.

\subsection{Clean Niobium Regime}

Niobium cavities in the clean regime are characterized by a very low concentration of impurities, i.e. very large residual resistivity ratio (RRR) and mean-free-path ($\ell$) within the RF penetration depth \cite{Padamsee_Book1}. To reach such a high level of purity, cavities are processed as follows. After their fabrication, approximately 120 $\mu m$ of material is removed from the inner cavity surface via electropolishing (EP) to eliminate the so-called damage layer; subsequently, they are heat-treated in an ultra-high-vacuum furnace at 800-900 \degree C for three hours to allow hydrogen degas from the bulk of the cavity and release stress in the material. To ensure that no contaminants diffuse from the furnace to the cavity surface, another 20 $\mu m$ is usually removed via EP. Some of the cavities processed and tested in this study underwent several cycles of surface treatments. This section reports the data acquired from cavities that were "reset" to the clean niobium regime by removing at least 40 $\mu m$ from the inner cavity surface via EP.
\begin{figure}[t]
\centering
\includegraphics[width=\columnwidth]{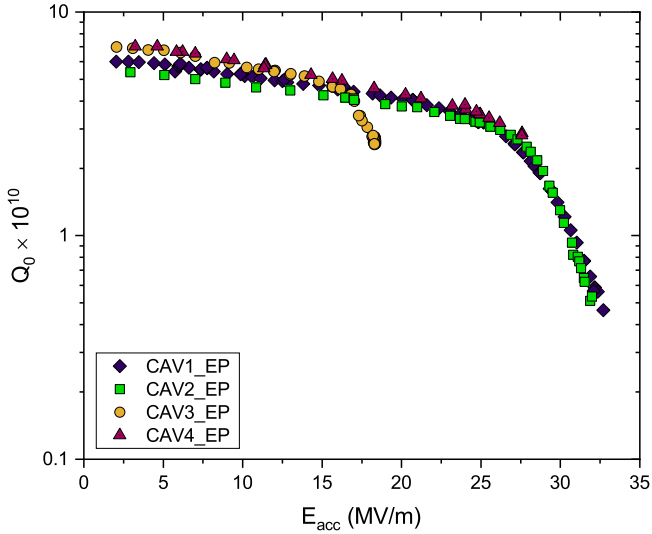}
\caption{Q-factor versus accelerating field data measured at 2.0 K for the single-cell EP cavities.}
\label{FigEP}
\end{figure}
\begin{figure}[t]
\centering
\includegraphics[width=\columnwidth]{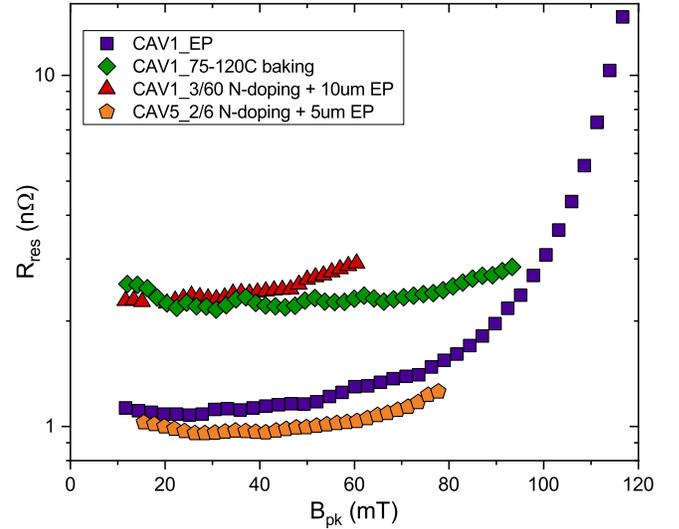}
\caption{Example of residual resistance as a function of the peak magnetic field for each of the surface treatments analyzed in this paper: EP, 75-120 \degree C baking, 2/6 and 3/60 N-doping.}
\label{Residual}
\end{figure}
\begin{figure}[t]
\centering
\includegraphics[width=\columnwidth]{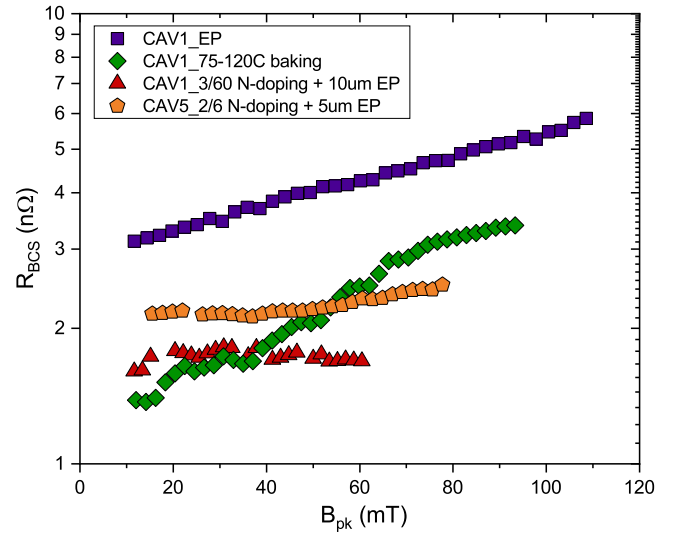}
\caption{Example of 2 K BCS surface resistance as a function of the peak magnetic field for each analyzed surface treatment in this paper: EP, 75-120 \degree C baking and 2/6 and 3/60 N doping.}
\label{BCS}
\end{figure}

In Fig. \ref{FigEP}, the Q-factor versus accelerating field measured at 2 K is shown for the studied cavities. Very high Q-factors that characterize these cavities at low RF fields ($Q_0=7-8\times 10^{10}$) are noticeable. In almost all curves, the Q-factor constantly decreases as a function of the field until approximately 25 MV/m, after which the high-field Q-slope (HFQS) prevails and rapidly decreases  the Q-factor as a function of the accelerating field. CAV3 shows HFQS-like behavior starting at a lower accelerating field, possibly due to defects in the cavity. Additionally, CAV3 and CAV4 showed premature quenching at approximately 17 and 27 MV/m, respectively. The other two cavities (CAV1 and CAV2) were not limited by the quenching of the superconducting state, but by the available RF power. Field emission was not observed in any test.

In Figs. \ref{Residual} and \ref{BCS} the residual resistance, $R_{\mathrm{res}}$, and BCS surface resistance, $R_{\mathrm{BCS}}$, are plotted against the peak RF magnetic field, $B_{\mathrm{pk}}$, respectively, for each of the surface treatments analyzed in this paper: EP, 75-120 \degree C baking, and 2/6 and 3/60 N-doping. For this cavity geometry the relation between the accelerating field and the peak magnetic field is the following: $B_{pk}/E_{acc}=3.888$.The surface resistance measured at 1.4-1.5 K is approximated as the residual resistance; therefore, the BCS contribution is estimated by subtracting the residual resistance from the surface resistance measured at 2.0 K. The $R_{\mathrm{BCS}}$ of EP cavities shows a linear increase as a function of the RF field within the entire analyzed range of fields, while $R_{\mathrm{res}}$ shows a linear trend until approximately 80 mT and starts to exponentially increase in larger RF fields. This behavior is typical of clean niobium cavities subjected to high-T baking and EP only. One interesting characteristic is that the onset of the exponential growth of the residual resistance responsible for the HFQS in the Q-factor versus accelerating field curve appears at approximately 80 mT, while for 1.3 GHz cavities it is usually observed at approximately 100 mT \cite{Romanenko_decomposition}.
A frequency dependence of the HFQS onset is not observed when examining a wider range of frequencies \cite{Ciovati_3GHz}. Therefore the different onset of the residual resistance exponential growth may be due, for example, to a different intake of hydrogen during the electropolishing treatment instead of being fundamentally linked to the frequency. This is consistent with the theory postulating that the HFQS phenomena are due to the incremental breaking of the proximity coupling of nanohydrides dissolved in the niobium near-surface region \cite{Romanenko_hydrides}. Because of the lack of significant statistics for this treatment, we cannot exclude the possibility of a local defect on the inner cavity surface. 

\subsection{Low-T baking treatments}
Very long baking at relatively low temperatures is known to be an effective "cure" for the high-field Q-slope and enables niobium cavities to reach an accelerating gradient near the theoretical limit imposed by the superheating field $H_{\mathrm{sh}}$ \cite{Hsp1, Hsp2, Hsp3}. Typically, after the cavity has been annealed and electropolished as described in the previous section, it is assembled and baked in situ at 120 \degree C for 48 hours \cite{Padamsee_Book2}. It was recently discovered that accelerating gradients might be further improved when in-situ low-T baking is first performed for 4 hours at 75 \degree C and subsequently for 48 hours at 120 \degree C \cite{75_120C_baking}. Therefore, this last surface treatment is a modified version of the standard 120 \degree C baking.

\begin{figure}[t]
\centering
\includegraphics[width=\columnwidth]{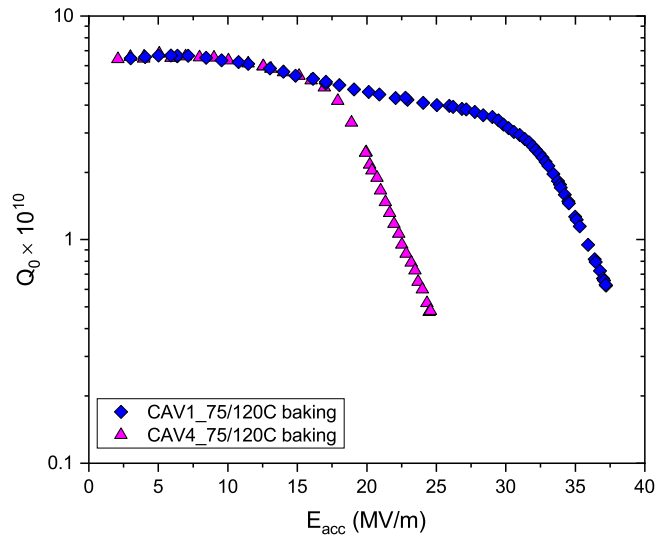}
\caption{Q-factor versus accelerating field data measured at 2.0 K for the single-cell low-T baked cavities.}
\label{FigLowT}
\end{figure}

In Fig. \ref{FigLowT} the Q-factor as a function of the accelerating field is shown for cavities that were subjected to the modified 120 \degree C baking treatment. Additionally, in this case, the Q-factor at low RF fields is very high: it starts from approximately $Q_0=7\times 10^{10}$ and slightly decreases when the RF field increases. In both analyzed cavities, the HFQS was not cured by low-T baking, and in the best case, it was only pushed to a higher RF field (approximately 29 MV/m, which corresponds to 113 mT). 
CAV4 shows the appearance of HFQS at approximately 18 MV/m ($\sim$ 70 mT), which is earlier than what was observed for clean EP cavities in the earlier section. The same cavity was, however, quenching prematurely at approximately 27 MV/m before being low-T baked (red curve in Fig. \ref{FigEP}). Therefore, low-T baking may have lowered the superconducting properties of defects already present on the surface of this cavity.
In both cases, after the modified 120 \degree C baking treatment, the measurements were not limited by the quenching of the superconducting state but by the available RF power.

Examples of the residual and BCS surface resistances of a low-T baked cavity are shown in Figs. \ref{Residual} and \ref{BCS}, respectively.
$R_{\mathrm{BCS}}$ starts from a very low value at a low RF field (1.5 n$\Omega$) and rapidly increases as a function of the RF field.
As expected from studies conducted in 1.3 GHz cavities \cite{Romanenko_decomposition}, $R_{\mathrm{res}}$ is larger than in clean niobium cavities at low and medium RF fields. With higher RF fields, the residual resistance of clean Nb cavities exponentially grows in the HFQS region.

\subsection{N doping for high Q at medium gradient}
N doping is applied to niobium cavities to obtain very high Q-factors at medium RF fields of approximately 15-20 MV/m \cite{Grassellino_SUST_2013}. However, as mentioned in the introduction, previous studies have shown that at these frequencies, N-doped cavities do not show the anti-Q-slope typically observed in 1.3 GHz cavities\cite{Martinello_SRF_2017, Martinello_PRL}.

N doping is performed immediately after high-T baking, when the cavity is still in the high-temperature furnace, by injecting 25 mTorr of nitrogen at 800 \degree C for some minutes. The process may be followed by a diffusion step, where nitrogen is shut off but the cavity remains at 800 \degree C for a certain amount of time before one turns off the temperature and lets the system cool naturally. After this doping process in the furnace, some micrometers must be removed via EP from the inner cavity surface to eliminate nonstoichiometric niobium nitrides that form during the process.

In this manuscript, we have studied the effects of two different N-doping treatments: the so-called "2/6" and "3/60" treatments \cite{Grassellino_SRF_2015, Palczeski_SRF_2019}. The first number indicates the amount of time (in minutes) for which the cavity is exposed to nitrogen in the furnace, while the second number indicates the amount of time (in minutes) of the diffusion step. We have also studied the effect of different post doping EP removal methods to understand how to maximize the Q-factor at medium RF fields and quench fields.  
\begin{figure}[t]
\centering
\includegraphics[width=\columnwidth]{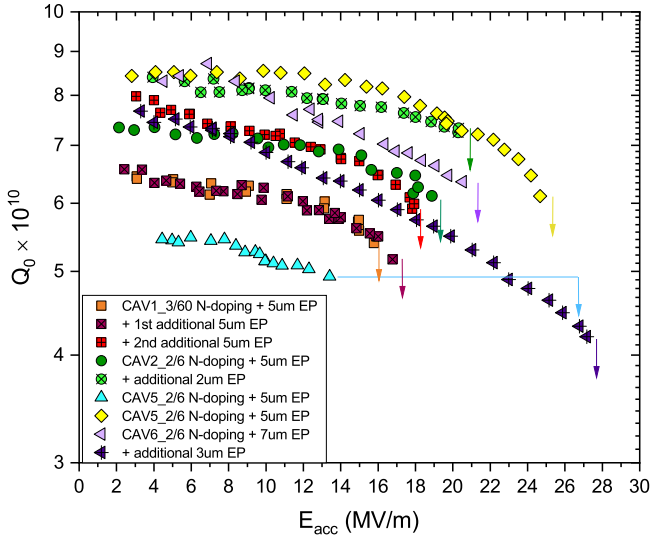}
\caption{Q-factor versus accelerating field data measured at 2.0 K for the single-cell N-doped cavities. The arrows indicate the quench field value for each cavity.}
\label{FigDoping}
\end{figure}

The Q-factor versus accelerating field curves of the analyzed N-doped cavities are shown in Fig. \ref{FigDoping}. Some spread in the Q-factor values is expected to be due to magnetic flux being trapped during some of the measurements. All cavities except CAV6 were tested in a compensated magnetic field by means of a couple of Helmholtz coils. This method usually guarantees a magnetic field lower than 1 mG around the cavity when cooled through its critical temperature. However, it is important to consider that, as observed for 1.3 GHz cavities, N-doped 650 MHz cavities show an increased sensitivity to trapped flux \cite{Martinello_APL_2016,Checchin_Sensitivity_Freq}. For 650 MHz N-doped cavities, it is expected that when 1 mG is trapped in the cavity, the surface resistance increases by approximately 1.5 n$\Omega$ at medium RF fields, causing a somewhat significant lowering of the Q-factor. Nonetheless, N-doped cavities undoubtedly reach the highest Q-factors at medium RF fields ($Q_0>5\times 10^{10}$) among the processes studied. 
The highest Q-factor values are obtained by applying 2/6 N doping plus an additional 7 $\mu$m of EP removal. Interestingly, the Q-factor versus accelerating field curve of the cavity treated with 3/60 N doping is identical after 5 $\mu$m and 10 $\mu$m of EP removal. This result indicates that the N-doping profile must be constant within 5 and 10 $\mu$m from the cavity surface.
In all of these cases, the maximum achievable accelerating field was limited by the cavity quench. The average quench field of N-doped cavities is $<E_{\mathrm{acc}}>=22 (4)$ MV/m, with an average of $<E_{\mathrm{acc}}>=24 (3)$ MV/m for 2/6 N-doped cavities and $<E_{\mathrm{acc}}>=17 (1)$ for 3/60 doped cavities. The quench field values of the 2/6 N doped cavities agree with those observed in the 9-cell LCLS-II production cavities \cite{Gonnella_statistics}. However, the 3/60 N doped cavities show a considerable reduction in quench field compared to the 1.3 GHz cavities \cite{Bafia_Ndop}.

In Fig. \ref{Residual} and \ref{BCS}, $R_{\mathrm{res}}$ and $R_{\mathrm{BCS}}$ as a function of $B_{\mathrm{pk}}$ are shown for the 2/6 and 3/60 N-doping treatments. 
$R_{\mathrm{BCS}}$ is lower for 3/60 than for 2/6, but on the other hand, $R_{\mathrm{res}}$ is considerably higher. Additionally, the $R_{\mathrm{BCS}}$ surface resistance is almost flat as a function of the field for the 3/60 doped cavity, while it slightly increases as a function of the field for the 2/6 doped cavity. These findings confirm that at this frequency, N doping is not effective in reversing the BCS surface resistance. 

\subsection{Discussion}
\begin{figure}[b]
\centering
\includegraphics[width=\columnwidth]{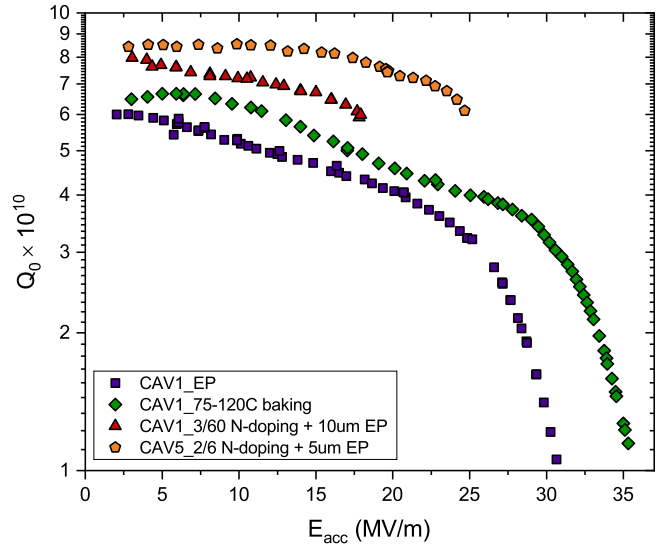}
\caption{Q-factor versus accelerating field measured at 2.0 K in single-cell cavities subjected to the surface treatment analyzed in this paper: EP, 75-120 \degree C baking, and 2/6 and 3/60 N doping.}
\label{CompQ}
\end{figure}

The single-cell study reveals some interesting findings, in particular: i) the HFQS onset of clean niobium cavities appears at $\sim$ 25 MV/m (97 mT) in agreement with higher frequency cavities, while the residual resistance exponential growth starts at $\sim$ 80 mT; ii) the HFQS cannot be mitigated with low-temperature baking, but its onset may be pushed to a higher RF field; iii) N-doped cavities show the highest Q-factors at medium RF fields. To strengthen this last point, Fig. \ref{CompQ} shows the comparison between the Q-factor versus accelerating field curves measured at 2.0 K in single-cell cavities subjected to the surface treatment analyzed in this paper: EP, 75-120 \degree C baking and 2/6 and 3/60 N doping. 

As discussed in the previous section, the observation of early residual resistance exponential growth is unlikely to be due to a real frequency dependence of the HFQS phenomenon. Meanwhile, this particular behavior resembles that typically observed in clean niobium cavities treated with buffer chemical polishing (BCP) instead of EP \cite{Ciovati_120Cbaking}. The cavity surface after BCP is rougher than that after EP \cite{BCP}, and a different hydrogen intake may also be expected. The similarity of the RF behaviors of these cavities suggests that the EP treatment may not be fully optimized for this cavity shape and promotes etching more than polishing. This hypothesis may also explain the early quenching observed in the 3/60 N-doped cavities. Recent studies conducted in 1.3 GHz 9-cell cavities have indeed shown that the quench field of N-doped cavities is improved by decreasing the temperature of the post-doping EP treatment \cite{Grassellino_SRF2019, Palczewski_SRF2019, Checchin_ColdEP}. A colder temperature helps slow the process and favors the polishing regime.

Our findings also confirmed that N doping maximizes Q-factors at medium RF fields in 650 MHz cavities. Even though the reversal of $R_{\mathrm{BCS}}$ as a function of the RF field is not observed, the presence of impurities still minimizes $R_{\mathrm{BCS}}$ between 60-80 mT, enabling very high Q-factors within the PIP-II operating gradient.
%
%
\section{N-doping optimization for multicell cavities}
Since N doping resulted in the best performance in terms of the Q-factor in single-cell cavities, we implemented the same treatment in the PIP-II $\beta = 0.9$ 5-cell 650 MHz cavities. 
\begin{figure}[b]
\centering
\includegraphics[width=\columnwidth]{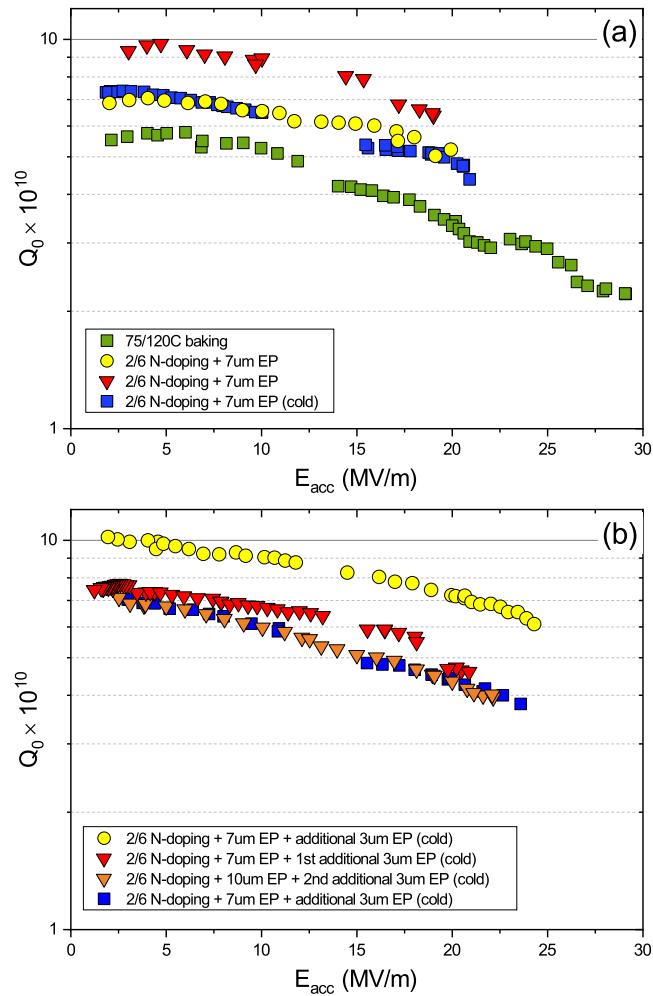}
\caption{Q-factor versus accelerating field measured at 2.0 K in 5-cell cavities subjected to [a] 2/6 N-doping followed by 7 $\mu$m EP and 75-120 \degree C baking and [b] 2/6 N-doping followed by 10 and 13 $\mu$m EP.}
\label{5-cell}
\end{figure}

Fig. \ref{5-cell} [a] shows the Q-factor versus accelerating field curves measured at 2.0 K in three N-doped cavities in comparison with that in a cavity treated with 75/120 \degree C baking. These results are consistent with our previous results in single-cell cavities: N doping enables us to obtain higher Q-factors at medium RF fields compared to the low-T treatment, but the quench field decreases to 19-21 MV/m. The highest quench field was achieved by the cavity that was subjected to cold EP after the N-doping treatment. This result is consistent with the previously mentioned hypothesis that cold EP helps promote cavity surface polishing.

\begin{figure}[t]
\centering
\includegraphics[width=\columnwidth]{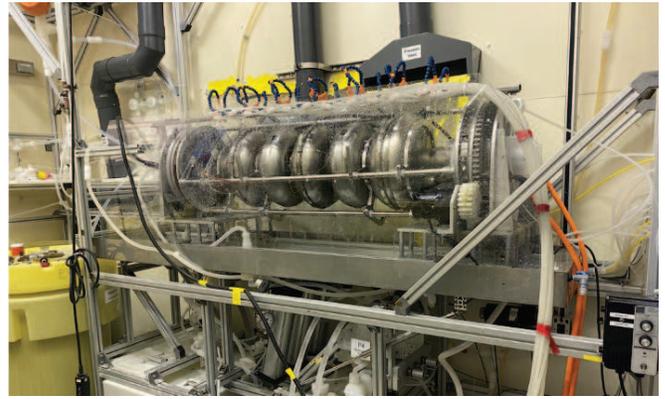}
\caption{Image of the setup at Argonne National Laboratory (ANL) to apply EP to the 650 MHz cavities.}
\label{EP}
\end{figure}

The electropolishing setup at Argonne National Laboratory (ANL) is shown in Fig. \ref{EP}, which enables cold EP to be performed in these 5-cell 650 MHz cavities. The external water spray system allows one to control the cavity surface temperature below 15 \degree C. The acid bath is also actively cooled during the process. This system is similar to the one at Fermilab which is utilized to perform cold EP in single-cell 1.3 GHz cavities \cite{Furuta_SRF_2019}.

To verify the optimal amount of post doping EP removal, the cavities were subjected to an additional 3 $\mu$m of cold EP for a total of 10 $\mu$m of EP removal after doping.  In Fig. \ref{5-cell} [b], the results of the 2.0 K measurements are summarized. In all cases, the quench field improved, and the cavity gradient was limited between 21-25 MV/m. One cavity was subjected to a second step of 3 $\mu$m of cold EP for a total of 13 $\mu$m of EP removal after doping. This second cold EP step helped the cavity to further push its quench field from 21 to 23 MV/m. 
Q-factors exceed $4\times 10^{10}$ at 20 MV/m in all cases, and some scattering in the Q-factor is most likely due to different magnetic field backgrounds during the vertical test measurement.

\section{Conclusions}
We investigated the effects of different state-of-the-art surface treatments on the performance of high-beta single-cell 650 MHz cavities. Our findings confirm that, even in this regime, N doping enables us to obtain the highest Q-factor values at medium RF fields.
The nature of the early quenching observed in these cavities must be better understood and is most likely related to a non optimal EP process.

The initial study in single-cell cavities enabled us to define the processing required to obtain high Q at medium RF fields in PIP-II 5-cell HB650 cavities. High-Q processing was then successfully transferred from single-cell cavities to PIP-II HB650 cavities, which qualified the first set of cavities for jacketing and subsequent cryomodule assembly.

Further studies that focus on extending this research in PIP-II LB650 cavities are currently ongoing. 

All of these studies are of crucial importance to advance both the performance and the fundamental understanding of elliptical SRF cavities with resonance frequencies lower than 1.3 GHz. These types of cavities are of great interest for future particle accelerators for both high-energy and nuclear physics. 

\section{Acknowledgments}

This work was supported by the United States Department of Energy, Offices of High Energy and Nuclear Physics. Fermilab is operated by Fermi Research Alliance, LLC under Contract No. DE-AC02-07CH11359 with the United States Department of Energy.

\section{Conflict of interest}

The authors have no conflicts to disclose.

\section{Data Availability}
The data that support the findings of this study are available from the corresponding author upon reasonable request.

\null 

\end{document}